\documentclass[reprint,superscriptaddress,amsmath,amssymb,aps,prb,floatfix,]{revtex4-2}
\usepackage{graphicx}% Include figure files
\usepackage{dcolumn}% Align table columns on decimal point
\usepackage{bm}% bold math
\usepackage{xcolor}
\usepackage{relsize}
\usepackage{amsmath}
\usepackage{amsfonts}
\usepackage{mathtools}
\usepackage{braket}
\usepackage{url}
\usepackage[colorlinks = true,
            linkcolor = blue,
            urlcolor  = blue,
            citecolor = blue,
            anchorcolor = blue]{hyperref}
\usepackage{footmisc}
\usepackage[utf8]{inputenc}
\usepackage[T1]{fontenc}

\begin{document}

\title{Collective excitations of exciton-polariton condensates in a synthetic gauge field}

\author{D.~Biega\'{n}ska}
\author{M.~Pieczarka}
\affiliation{ARC Centre of Excellence in Future Low-Energy Electronics Technologies and Nonlinear Physics Centre, Research School of Physics, The Australian National University, Canberra, ACT 2601, Australia}
\affiliation{Department of Experimental Physics, Wroc\l aw University of Science and Technology, Wyb. Wyspia\'{n}skiego 27, 50-370 Wroc\l aw, Poland}%
\author{E.~Estrecho}
\affiliation{ARC Centre of Excellence in Future Low-Energy Electronics Technologies and Nonlinear Physics Centre, Research School of Physics, The Australian National University, Canberra, ACT 2601, Australia}
\author{M.~Steger}
\thanks{Current address: National Renewable Energy Lab, Golden,
CO 80401, USA}
\author{D.~W.~Snoke}
\affiliation{Department of Physics and Astronomy, University of Pittsburgh, Pittsburgh, PA 15260, USA}
\author{K.~West}
\author{L.~N.~Pfeiffer}
\affiliation{Department of Electrical Engineering, Princeton University, Princeton, NJ 08544, USA}
\author{M.~Syperek}
\affiliation{Department of Experimental Physics, Wroc\l aw University of Science and Technology, Wyb. Wyspia\'{n}skiego 27, 50-370 Wroc\l aw, Poland}
\author{A.~G.~Truscott}
\affiliation{Laser Physics Centre, Research School of Physics, The Australian National University, Canberra, ACT 2601, Australia}
\author{E.~A.~Ostrovskaya}
\email{elena.ostrovskaya@anu.edu.au}
\affiliation{ARC Centre of Excellence in Future Low-Energy Electronics Technologies and Nonlinear Physics Centre, Research School of Physics, The Australian National University, Canberra, ACT 2601, Australia}
\date{\today}

\begin{abstract}
Collective (elementary) excitations of quantum bosonic condensates, including condensates of exciton polaritons in semiconductor microcavities, are a sensitive probe of interparticle interactions. In anisotropic microcavities with momentum-dependent TE-TM splitting of the optical modes, the excitations dispersions are predicted to be strongly anisotropic, which is a consequence of the synthetic magnetic gauge field of the cavity, as well as the interplay between different interaction strengths for polaritons in the singlet and triplet spin configurations. Here, by directly measuring the dispersion of the collective excitations in a high-density optically trapped exciton-polariton condensate, we observe excellent agreement with the theoretical predictions for spinor polariton excitations. We extract the inter- and intra-spin polariton interaction constants and map out the characteristic spin textures in an interacting spinor condensate of exciton polaritons.
\end{abstract}

\maketitle

\textit{Introduction --} Exciton polaritons (polaritons herein) are quasiparticles formed by excitons strongly coupled to confined photons, typically in a semiconductor optical microcavity. As interacting bosons, they form 2D nonequilibrium condensates analogous to Bose-Einstein condensates (BECs) of ultracold atoms at sufficiently large particle densities above the phase transition threshold \cite{KavokinMICROCAV,Sanvitto2010,Amo2009,Kasprzak2006}. Furthermore, polaritons possess a spin degree of freedom inherited from optically active excitons coupled to photons \cite{Shelykh_2009,Carusotto2013,Vladimirova2010}. Polariton spin has two allowed integer projections on the cavity growth axis, $\sigma_{\pm}$, making a polariton condensate effectively a two-component (spinor) gas described by a pseudospin parameter \cite{Pitaevskii2016,Kavokin_microcavities2008}. 

Polaritons interact through their excitonic components \cite{Carusotto2013,Deveaud-Pledran2016}, with a spin-dependent strength \cite{Renucci2005,Vladimirova2010,Takemura2014,bleu2020polariton}. The interaction strengths for polaritons of the same (triplet) $\alpha_{1}$ and opposite (singlet) $\alpha_{2}$ spin are related as $|\alpha_2 |\ll\alpha_1$ \cite{Vladimirova2010,Deveaud-Pledran2016,Sekretenko2013}. On the other hand, momentum-dependent TE-TM splitting of the optical modes of the microcavity \cite{Panzarini1999} and optical anisotropy (linear birefringence) of the cavity create an effective magnetic field, which affects the polariton dynamics in the low-density regime, below the condensation threshold, via the photonic component of the quasiparticles \cite{Leyder2007,Kavokin2005OSHE}, similar to other optical systems \cite{Rechcinska2019,Bliokh2015}. In the high-density regime, above the condensation threshold, this synthetic field affects the condensate pseudospin dynamics \cite{Kammann2012,Hivet2012,Cilibrizzi2016,Schmidt2017} in addition to the effect of spinor polariton-polariton interactions. As a result, the single-particle dispersion of the polaritons in the low-density regime, as well as the dispersion of the collective excitations \cite{Utsunomiya2008} of the condensate are predicted to be strongly anisotropic \cite{ShelykhPRL2006,TercasPRL2014,SOLNYSHKOV2016Chirality}. Namely, the dispersion branches cross in one of the directions in the 2D momentum space at the so-called diabolical points, forming the characteristic Dirac cones. The characteristic monopolar pseudospin texture around these crossing points in momentum space can be described in terms of an effective Rashba-like non-Abelian gauge field \cite{TercasPRL2014,SOLNYSHKOV2016Chirality,Gianfrate2020}. Studies of such gauge fields were previously limited to utracold atomic BECs \cite{Lin2011}. Observation of a synthetic (artificial) gauge field for polaritons in anisotropic microcavities offers the possibilty to study topological phases of matter \cite{Hasan2010,Klembt2018,Bardyn2015} and analogue physics in optical systems \cite{Fieramosca2019}. However, despite the experimental progress in mapping the nontrivial spin textures in the single-particle (linear) regime \cite{Gianfrate2020}, the predicted behaviour of collective excitations in the interactions-dominated (nonlinear) regime above the condensation threshold has not been confirmed so far \cite{Kohnle2011,Pieczarka2015,Stepanov2019,Ballarini2020}.

In this Letter, we observe the dispersion of collective excitations of a linearly polarized high-density polariton condensate in an optical trap. By performing polarization-resolved photoluminescence tomography, we detect the excitation branches in momentum space and observe a clear asymmetry in directions parallel and perpendicular to the cavity anisotropy axis. Moreover, we determine the triplet and singlet interactions strengths $\alpha_{1,2}$ and extract the spin textures and synthetic magnetic gauge field distribution in the nonlinear regime.

\textit{Experimental setup --}
We use an ultrahigh-quality GaAs-based microcavity cooled down to $\sim$4K using a continuous-flow liquid helium cryostat. The very narrow linewidth of the polariton emission arising from the long cavity photon lifetime of $>100\:$~ps in this sample \cite{Steger2015} enables us to resolve the non-negligible anisotropy of the polariton dispersion~\cite{Gianfrate2020}. Off-resonant excitation of the sample with a continuous-wave laser beam, shaped by a conical lens into a ring of $45~\mu\text{m}$ in diameter, creates a round ``box" optical trap~\cite{Estrecho2019, Pieczarka2019} for polaritons. This trapping geometry minimizes the overlap of polaritons with the excitonic reservoir. We record the photoluminescence spectra in 2D momentum space by translating the imaging lens with respect to the monochromator slit. The polarization sensitivity is achieved by employing a half-waveplate and a linear polarizer in the detection path, enabling us to record the spectra in four linear polarizations bases: horizontal/vertical (H/V) and diagonal/antidiagonal (D/A) with respect to the laboratory frame of reference. The experiments are performed on a region of the sample corresponding to small, positive exciton-photon detuning of $\Delta=(2.70\pm0.21)~\text{meV}$ and the excitonic Hopfield coefficient $X_0^2\approx 0.585$, which defines the excitonic fraction of the polariton. Further details of the experiment can be found in the Supplemental Material.

\begin{figure}[t]
    \centering
    \includegraphics[width=\columnwidth]{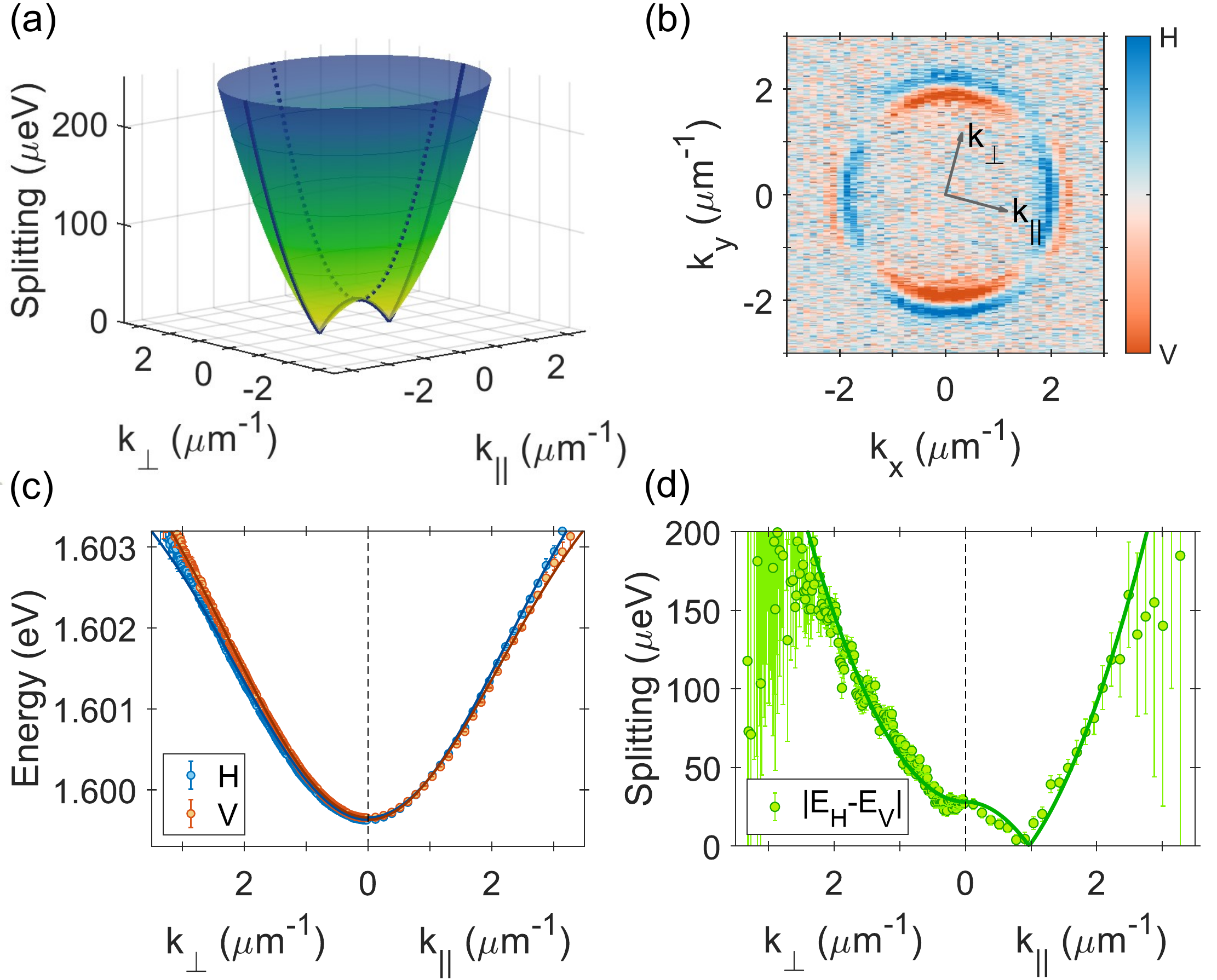}
    \caption{Polariton dispersion in the low-density regime. (a) Energy splitting between the polariton eigenstates, calculated from Eq.(\ref{eq:H_sp}), with the cross-sections in the directions parallel (solid line) and perpendicular (dotted line) to the anisotropy axis. (b) Cross-section of the Stokes vector component $S_1$ at the constant energy $E=1.60154~\text{eV}$ in the measurement ($k_x,k_y$) and anisotropy ($k_{||},k_\perp$) frames of reference. (c) Polariton eigenstates and (d) their energy splitting extracted from the measured dispersions in the directions $k_{||}$ and $k_{\perp}$. Solid lines are the model fits.}
    \label{fig1}
\end{figure}

\begin{figure*}[ht]
    \centering
    \includegraphics[width=\textwidth]{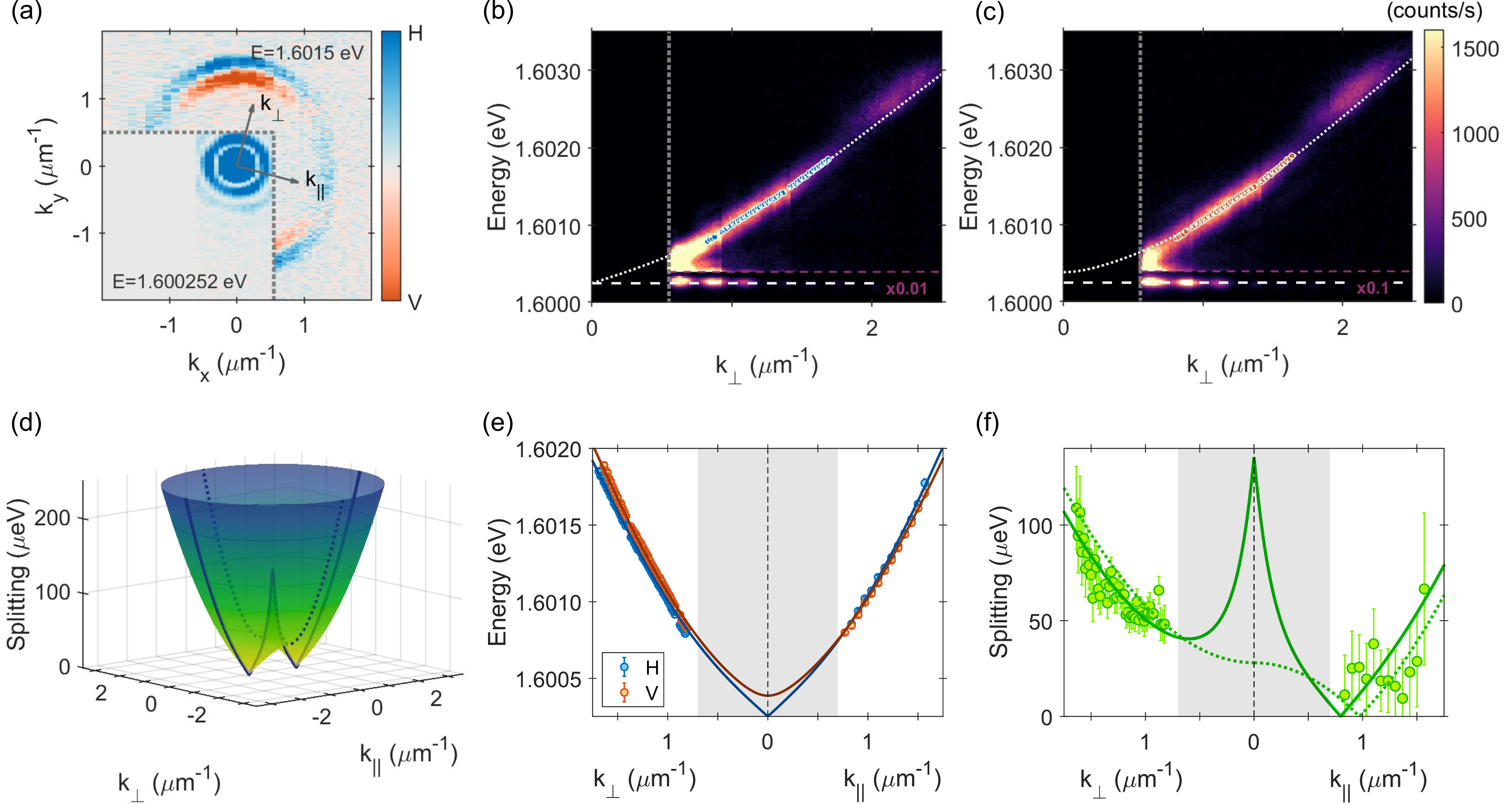}
    \caption{Dispersion of collective excitations in the high-density regime. (a) Cross-section of the Stokes vector component $S_1$ at the constant energy $E = 1.60154~\text{eV}$. Inset: horizontally polarized condensate at $E=\mu =1.600252~\text{eV}$, with visible Airy pattern due to the spatial filter (see text). Grey dotted lines mark the edges of the filtered area. The frames of reference are marked as in Fig.~\ref{fig1}(b). (b,c) Examples of the photoluminescence spectra at $k_\parallel=0$ in (b) horizontal (H) and (c) vertical (V) polarization basis. Data points show energies extracted from fitting, white dotted lines - the fitted model. White dashed line marks the energy of the condensate and grey dotted line is the edge of the spatial filter in momentum space. The condensate emission in the image is attenuated by a factor of 0.01 (b) and 0.1 (c). (d) Energy splitting between the excitations predicted by the model with the cross-sections in the directions parallel (solid line) and perpendicular (dotted line) to the anisotropy axis. (e) Dispersions and (f) energy splitting between the collective excitations in the directions parallel and perpendicular to the condensate polarization direction with the corresponding theoretical fits (solid lines). Green dashed line in (f) shows the energy splitting for the single-particle states. Grey shaded area in (e,f) marks the range of emission blocked by the edge filter.}
    \label{fig2}
\end{figure*}

\textit{Low-density regime --} 
Polariton eigenstates in the low-density limit can be described by a single-particle Hamiltonian in a circular polarization basis~\cite{TercasPRL2014,Gianfrate2020}:
\begin{equation}
\label{eq:H_sp}
    H = 
    \begin{pmatrix}
    \epsilon_{LP}(k) & \frac{\Omega}{2}e^{-i\varphi}-\beta k^2e^{-2i\theta_k} \\
    \frac{\Omega}{2}e^{i\varphi}-\beta k^2e^{2i\theta_k} & \epsilon_{LP}(k) \\
    \end{pmatrix},
\end{equation}
where $\epsilon_{LP}$ is the lower-polariton dispersion extracted from a coupled-oscillator model,  $\beta$ is the TE-TM splitting parameter, and $\Omega$ is the cavity anisotropy constant. The wavevector can be expressed as $\mathbf{k}=k(\cos{\theta_k},\sin{\theta_k})$, with $\theta_k$ denoting the in-plane propagation angle. The angle $\varphi$ defines the anisotropy axis that depends on the sample orientation. By diagonalizing the Hamiltonian, one obtains two dispersive, linearly polarized eigenstates. The cavity anisotropy, $\Omega$, breaks the cylindrical symmetry of the TE-TM splitting resulting in both energy and polarization splitting at $k=0$, as shown in Fig.~\ref{fig1}(a,c,d). The two dispersion branches diverge in the direction perpendicular to the anisotropy axis ($k_{\perp}$), but cross in the direction parallel to it ($k_{\parallel}$). The crossing point occurs at $k_{\parallel}^*=\sqrt{\Omega/(2\beta)}$, where the effects due to TE-TM splitting and optical anisotropy cancel each other [Fig.~\ref{fig1}(d)].

To apply this model to the experimental data, we fit the polarization-resolved spectra for each wavevector with a Lorentzian function and extract the energy of the eigenstate from the spectral peak. Subsequently, by fitting the measured dispersions with the eigenvalues of Eq.~\ref{eq:H_sp}, as shown in Fig.~\ref{fig1}(c),  we find the values: $\beta=(14.89\pm0.92) \:\mu eV \:\mu m ^{-2}$, $\Omega=(28.1\pm2.2) \:\mu eV $, and $\varphi=-15^\circ$. Hence, the crossing point occurs at $k_{\parallel}^*=0.971 \:\mu m^{-1}$ and $k_{\parallel}$ is $-15^\circ$ from the $+x$-axis (see the orientation of the frames of reference in Fig.~\ref{fig1}(b)). These measured parameters are essential for the analysis of the collective excitations in the high-density regime.

For each of the eigenstates, we also extract two components of the Stokes vector, $S_1,S_2$, which correspond to the polarization state or in-plane pseudospin of the polaritons. They are calculated from the polarized photoluminescence intensities, $I$, using the formulas: $S_1=(I_H-I_V)/(I_H+I_V)$ and $S_2=(I_D-I_A)(I_D+I_A)$. An example of the extracted texture of the $S_1$ component is presented in Fig.~\ref{fig1}(b) and is consistent with the previous measurements in high-quality GaAs-based microcavities~\cite{Gianfrate2020}. This is a typical texture arising from TE-TM splitting, which is the dominant effect at large $k>k_{\parallel}^*$. The full in-plane pseudospin textures of the single-particle eigenstates are shown in the Supplemental Material.

\begin{figure*}
    \centering
    \includegraphics[width=\textwidth]{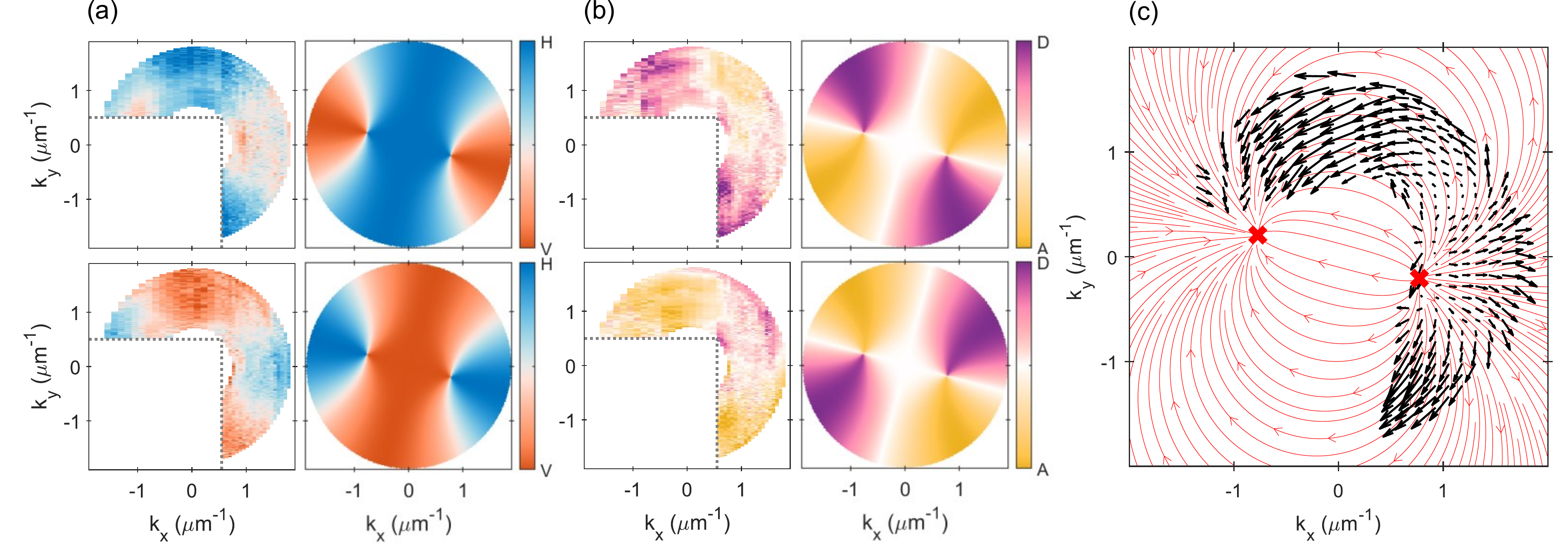}
    \caption{(a) $S_1$ and (b) $S_2$ components of the Stokes vector for the collective excitations. Shown are experimental (left column) and theoretical (right column) results for the lower excitation branch $\epsilon^L$ (top row) and the upper branch $\epsilon^U$ (bottom row). Color scales are normalized to the maximum of the Stokes parameter, where the maximum Stokes amplitudes of the experimental results are around $40\%$. (c) Calculated pseudospin texture of $\epsilon^U$ (black arrows), plotted over the field texture obtained from the theoretical model. Red crosses mark the positions of the crossing points.}
    \label{fig3}
\end{figure*}

\textit{Condensate excitations --} By increasing the pump power above the condensation threshold, we create a high-density, single-mode condensate in the Thomas-Fermi regime~\cite{Estrecho2019,Pieczarka2020}, formed in an optically induced potential trap. The condensate emission is highly polarized, with $95\%$ degree of linear polarization oriented parallel to the anisotropy angle $\varphi=-15^\circ$. A similar pinning of the condensate polarization to a given direction in the sample is routinely observed in different types of samples~\cite{Kasprzak2007,Balili2007,Levrat2010,Klaas2019,Gnusov2020} and is not a signature of the expected stochastic buildup of polarization due to spontaneous symmetry breaking~\cite{Ohadi2012}. Here, we provide a strong evidence that the pinning of the condensate polarization arises from the optical anisotropy of the cavity. This effect is undetectable in low-quality samples, where the anisotropic splitting $\Omega$ is much smaller than the spectral linewidths.

To study dispersion of the condensate excitations, we filter out the strong contribution of the condensate near $k=0$ with an edge filter in momentum space. This allows us to detect the much weaker emission from the excitations without saturating the camera~\cite{Pieczarka2020}. We use a vertically (horizontally) oriented edge to block the emission from $k_x< 0.55\:\mu\mathrm{m}^{-1}$ ($k_y< 0.5\:\mu\mathrm{m}^{-1}$). Tomographic scans are performed for each edge filter orientation and combined to reconstruct a 2D excitation spectrum. A constant-energy slice of the scans is presented in Fig.~\ref{fig2}(a) showing the momentum-resolved $S_1$ texture and the edges (dashed lines) of the filter. A circular real-space aperture is also used to block the photoluminescence from the annular barrier of the optical trap.

The images of the collective excitation branches along $k_\perp$ at $k_\parallel=0$ for H- and V-polarizations are shown in Fig.~\ref{fig2}(b,c). The residual emission at the condensate energy has the characteristic Airy pattern arising from the diffraction on the real-space filter \cite{Pieczarka2020}. Above the condensate energy, clear Bogoliubov excitation branches are seen in both orthogonal polarizations. The negative or ghost branches \cite{Pieczarka2020} are also detectable, but are extremely weak, therefore we focus on the much brighter normal branches in our detailed analysis. At a constant energy slice, see Fig.~\ref{fig2}(a), the two branches have different polarizations and wavevectors. Equivalently, the dispersions are split as shown in Fig.~\ref{fig2}(e). As predicted by theory and shown in Fig.~\ref{fig2}(d), the energy splitting is highly anisotropic, which is confirmed by the experimental data in Fig.~\ref{fig2}(f). Similarly to the low-density limit (Fig.~\ref{fig1}), the collective excitation branches diverge in the direction perpendicular to the anisotropy axis (along $k_\perp$) and cross in the orthogonal direction (along $k_\parallel$).

The experimental results can be modeled by solving the linearized equations for excitation eigenstates within the mean-field framework, as described in Ref.~\cite{TercasPRL2014}. Taking into account the optical anisotropy and the TE-TM splitting, the excitations of a linearly polarized condensate at $k=0$ result in four dispersion branches $\pm\epsilon^{L,U}$ -- two for positive and two for negative energies with respect to the condensate energy (chemical potential) $E=\mu$. 

%{\color{red} (should we write the matrix instead? This would be important for the polarization modes)}
%{\color{red} - Dabrówka: I find the exact solutions much clearer, but please all decide which should we use (maybe actually showing both would be best? Or too much space?), I include the full model simplified for our case and a short note} 
%{\color{blue}M: Put the full matrix to the supplementary material with a short description of obtaining eigenvalues.} 
%{\color{red} - Dabrówka: Following Maciej's advise I've put the full matrix into SM} 

Expressed in the basis aligned to the cavity anisotropy axis, the positive excitation branches can be written as:
\begin{equation}
\begin{split}
    \epsilon^U \left(q\right)= \\
    \sqrt{\left( \Omega + \epsilon(q) \mp \beta q^2\right)  \left( \Omega + \epsilon(q) \mp \beta q^2 + 2\left( \alpha_1 - \alpha_2\right) n\right)}
\end{split}
\end{equation}
\begin{equation*}
    \epsilon^L \left(q\right) = \sqrt{\left( \epsilon(q) \pm \beta q^2\right)  \left( \epsilon(q) \pm \beta q^2 + 2\left( \alpha_1 + \alpha_2\right) n\right)},
\end{equation*}
where $\epsilon(q) = \epsilon_{LP}(q) - \epsilon_{LP}(0)$, and $n_{tot}=2n$ is the total condensate density. Here the wavevector is aligned to the anisotropy axis such that ${q} = k_{\parallel,\perp}$, %{\color{blue}M: I bring back the previous description with "q" and a bit modified.}
and the order of the $\pm$ and $\mp$ signs corresponds to $k_\parallel$ and $k_\perp$. The two branches inherit the anisotropic behavior of the single-particle dispersions, as shown by the energy splitting in $k$-space [see Fig.~\ref{fig2}(d)]. However, the crossing points $k^*$ now also depend on the spin-dependent interaction constants.

Since the parameters $\beta$ and $\Omega$ are known from the low-density measurements, we fit the measured dispersion of excitations using $\alpha_1n$ and $\alpha_2n$ as fitting parameters. Good agreement between the experimental data and the theoretical model is illustrated in Fig.~\ref{fig2}(e,f). One can observe a shift of the crossing points $k^*$ with respect to the low-density case, directly induced by the polariton-polariton interactions in the condensed state. The resulting values are $\alpha_1 n=(322\pm12) \:\mu eV$ for the polaritons in the triplet spin configuration and $\alpha_2 n=(9\pm15) \:\mu eV$ in the singlet configuration, with the corresponding chemical potential $\mu=\left( \alpha_1+\alpha_2 \right) n -\frac{\Omega}{2}=(317\pm20) \:\mu eV$. The condensate density is measured \cite{Pieczarka2020} to be $\sim2200 \:\mu m^{-2}$. This yields the interaction constants $\alpha_1=(0.293\pm0.029) \:\mu eV \:\mu m^{2}$, while $\alpha_2=(0.008\pm0.014) \:\mu eV \:\mu m ^{2}$. 
These measured values are in good agreement with previous estimates. For polaritons in GaAs-based samples, it is common to neglect the singlet contribution to the total blueshift. This assumption is fully supported by our result, with $\alpha_{2}$ around two orders of magnitude smaller than $\alpha_{1}$. Our values yield the ratio ${\alpha_2}/{\alpha_1}\approx 0.03$, being positive and smaller than the common assumptions \cite{Vladimirova2010,Deveaud-Pledran2016} for the GaAs system. It falls into the known region of stability of phase space for a linearly polarized condensate \cite{bleu2020polariton,Vladimirova2010}. In contrast to previous reports, our approach enables a direct measurement of both values. 

%Moreover, the optical anisotropy $\Omega$ provides the strongest contribution to the anisotropy of the collective excitation branches in our system. This finding can guide future experiments on tailoring the anisotropy of excitations, since the optical birefringence of the cavity is much easier to manipulate than the strength of interaction between polaritons.

%\vspace{12pt}
\textit{Spin texture of excitations --} The in-plane pseudospin of the collective excitations is characterised by the Stokes vector for the excitation eigenstates.
Extracted polarization patterns are presented in the left columns of Fig.~\ref{fig3}(a,b), with the right column showing the corresponding solution of the full theoretical model. To highlight the contrast in the patterns in spite of a very small splitting between the branches, we calculate the maximum Stokes vector components at the energies slightly offset from the eigenstates (the details of the analysis can be found in Supplemental Material). The experimental results show a good correspondence to the polarization patterns of the eigenstates expected from the model. The resulting pseudospin textures [Fig.~\ref{fig3}(c)] show a pattern similar to that reported in the single-particle case \cite{Gianfrate2020}. However, the existence and position of the diabolical points with the associated effective monopolar magnetic field is now governed not only by the ratio of $\Omega$ and $\beta$, but also by the polariton interactions, since they arise from the collective (Bogoliubov) excitations of an interacting condensate. The exact crossing points and the corresponding monopoles in the pseudospin patterns are not directly accessible in our experiment, being too close to the strong condensate emission. Nevertheless, a clear manifestation of such a gauge field texture is visible in the experimental data [Fig.~\ref{fig3}(c)]. 

%\vspace{12pt}
\textit{Conclusions --}
In this paper, we have demonstrated anisotropy of collective excitations in a spinor exciton-polariton condensate, which results from the innate spin-anisotropy of polariton interaction and the optical anisotropy (birefrigence) of the microcavity under study. 
%The difference in the interactions between same-spin and opposite-spin polaritons has a smaller contribution to the anisotropy of excitations in a GaAs-based system than the optical anisotropy of the cavity. 
 The optical anisotropy provides the strongest contribution to the anisotropy of the collective excitation branches in our GaAs-based system. Our experimental method enables a new, direct measurement of the interaction constants $\alpha_{1}$ and $\alpha_{2}$ for the polaritons in the triplet and singlet spin configurations, and can be applied to other systems with different interaction strength ratios. In our sample, we confirmed a two orders of magnitude difference in the interaction strengths $\alpha_{1,2}$. 

Furthermore, we extracted the pseudospin textures of the collective excitations in the polariton condensate resulting from the interplay between the effective magnetic field of the microcavity and spin-dependent interactions. The presence of diabolical points with the associated spin structure characteristic of a monopole-like magnetic field signifies that polariton systems in the high-density (nonlinear) regime can be used in future studies of synthetic gauge fields and topological physics. The dominant role of the cavity birefrigence in the anisotropy of collective excitations points to a straightforward way to design synthetic gauge fields for quantum liquids of light by tailoring the optical anisotropy of microcavities.

\bibliography{references}

\end{document}